\documentclass[a4paper]{jpconf}
\usepackage{graphicx}

\usepackage[caption = false]{subfig}
\usepackage{graphicx}
\usepackage{dsfont}
\usepackage{amssymb,amsmath,amsfonts}
\newcommand{\re}{{\rm e}}
\newcommand{\ri}{{\rm i}}
\newcommand{\rd}{{\rm d}}
\newcommand{\half}{\mbox{$\textstyle \frac{1}{2}$}}

\begin{document}

\title{PT symmetry and the evolution speed in open quantum systems
\footnote{This paper is based on the talk presented at the PHHQP 
Seminar Series.}}

\author{Dorje C Brody$^{1,2}$}

\address{$^1$Department of Mathematics, University of Surrey, 
Guildford GU2 7XH, UK
\vspace{0.1cm}\\
$^2$St Petersburg National Research University of Information 
Technologies, Mechanics and Optics, St Petersburg 197101, Russia }

\ead{d.brody@surrey.ac.uk}

\begin{abstract}
The dynamics of an open quantum system with balanced gain and loss is not described by a PT-symmetric Hamiltonian but rather by Lindblad operators. Nevertheless the phenomenon of PT-symmetry breaking and the impact of exceptional points can be observed in the Lindbladean dynamics. Here we briefly review the development of PT symmetry in quantum mechanics, and the characterisation of PT-symmetry breaking in open quantum systems in terms of the behaviour of the speed of evolution of the state. 
\end{abstract}

\section{PT symmetry and quantum physics} 

We begin with a brief account of the development of PT 
symmetry in quantum mechanics. In 1998 Bender and Boettcher found 
that there is a family of Hamiltonians of the form ${\hat H}=\frac{1}{2}
{\hat p}^2 + {\hat x}^2(\ri{\hat x})^\epsilon$ that possesses real, positive, 
and discrete  eigenvalues for all $\epsilon\geq0$ \cite{Bender1}. These 
Hamiltonians, 
while not Hermitian, are nevertheless invariant under the action of 
parity-and-time (PT) reversal: 
${\hat p}\to{\hat p}$, ${\hat x}\to-{\hat x}$, and $\ri\to-\ri$. This observation 
led them to conjecture that generic PT-symmetric Hamiltonians may 
possess entirely real eigenvalues. Subsequently, eigenvalues of many 
other PT-symmetric Hamiltonians were investigated by numerous authors. 

A natural question thus arising is whether a consistent quantum theory can 
be developed 
by use of a PT-symmetric Hamiltonian. In standard quantum mechanics the 
reality of observables are ensured by the Hermiticity condition on the Hilbert 
space endowed with a Hermitian inner product. While Hermiticity may be 
replaced by the condition of PT symmetry to enforce the reality of observables, 
a Hilbert space endowed with a PT-conjugation inner product is problematic, 
for, the parity operator is trace free and squares to the identity so that half of its 
eigenvalues are negative. In other words, such Hilbert spaces are of the 
Pontryagin-type that come equipped with indefinite metrics \cite{pontryagin}. 
Because 
Hilbert-space norms are related to probabilities in quantum mechanics, 
negative norms would imply negative probabilities, which makes the theory 
unphysical. 

In 2002, however, Mostafazadeh \cite{Mostafazadeh}, and independently 
Bender \textit{at al}. \cite{Bender6} observed 
that for a PT-symmetric Hamiltonian ${\hat H}$ there exists another 
PT-symmetric observable that commutes with ${\hat H}$ such that its 
eigenvalues are either plus or minus one, depending on the parity type of 
the eigenfunctions of ${\hat H}$. Such an operator naturally has the 
interpretation analogous to a charge (C) operator. This means that a 
PT-symmetric 
Hamiltonian is automatically CPT-symmetric, so in fact we can speak of 
CPT-symmetric quantum theory for which the Hilbert-space inner product 
is defined with respect to a positive-definite CPT conjugation. In particular, 
writing ${\hat g}$ for the operator associated with the CP conjugation, 
the inner product of a pair of elements $|\psi\rangle$ and $|\varphi\rangle$ 
is given by $\langle\psi,\varphi\rangle=\langle\psi|{\hat g}|\varphi\rangle$. 
It can then be shown that 
the operator ${\hat g}$ is Hermitian and positive, so that it admits the 
interpretation of defining a metric on Hilbert space, and thus is sometimes 
referred to as the metric operator. 

Because ${\hat g}$ is Hermitian and positive, there is an operator ${\hat u}$, 
not unique, such that we can write ${\hat g}={\hat u}{\hat u}^\dagger$. Then 
the operator ${\hat u}$ can be used to map between the Hilbert space endowed 
with a standard Hermitian inner product and that endowed with a CPT inner 
product, because $\langle\psi,\varphi\rangle=\langle{\hat u}^\dagger\psi|
{\hat u}^\dagger\varphi\rangle$. 
Further, the operator ${\hat u}$ can be used to define a similarity 
transformation between a PT-symmetric observable and a Hermitian 
observable; whereas ${\hat g}$ can be used to determine the Hermitian 
adjoint of an observable ${\hat F}$ in the sense that ${\hat F}^\dagger=
{\hat g}^{-1}{\hat F}{\hat g}$. An operator satisfying such a condition is 
sometimes referred to as being `pseudo-Hermitian'. Once these 
structures associated with PT symmetry were uncovered, it became 
appreciated that the notion of pseudo-Hermiticity in itself has been around 
for some time \cite{Mackey}, and likewise the idea of defining Hilbert space 
inner product in terms of a nontrivial metric operator for characterising 
quantum theory has been proposed previously \cite{SGH}, albeit in a different 
context. 

The fact that there is a similarity transformation relating a PT-symmetric 
Hamiltonian to a Hermitian one naturally raises the question whether a 
quantum system 
described by a PT-symmetric Hamiltonian is in fact equivalent to standard 
Hermitian quantum mechanics. That they are indeed equivalent on account 
of the existence of a similarity transformation is the point that has been 
argued for long by Mostafazadeh \cite{Mostafazadeh2}, although in infinite 
dimensions one has to be cautious in making such an assertion. 

To discuss the question of equivalence it will be useful first to consider quantum 
systems modelled on finite-dimensional Hilbert spaces. In this case, all operators 
are bounded, so mathematical subtleties associated with unbounded operators 
do not enter the discussion. It is then tempting to conclude that a 
system described by a PT-symmetric Hamiltonian is just an alternative 
representation of the same system described by a Hermitian Hamiltonian, i.e. 
they are equivalent. This `equivalence' argument by itself does not quite 
account for 
exactly what goes on for two reasons: First, a PT-symmetric Hamiltonian 
admits more parametric degrees of freedom than a Hermitian Hamiltonian. For 
instance, in the case of a two-level system, a Hermitian Hamiltonian has four 
independent parameters, while a PT-symmetric Hamiltonian can have six 
independent parameters. More generally, for an $n$-level system a 
PT-symmetric Hamiltonian can have $n(2n-1)$ independent parameters, 
as opposed to $n^2$ for a Hermitian Hamiltonian. 
Because parameters in a Hamiltonian represent 
experimental setup, a blind statement of equivalence does not explain this 
discrepancy. Second, for a PT-symmetric Hamiltonian it is possible to vary 
parameters in the Hamiltonian to effect a phase transition for which in one 
phase the eigenstates of the Hamiltonian are PT symmetric (the unbroken 
phase), while in the other, broken phase they are not. In the broken phase, the 
eigenvalues of the Hamiltonian are no longer real---some, or all of them, come 
in complex conjugate pairs. No such phenomenon is seen in a Hermitian 
Hamiltonian in finite dimensions. 

This apparent contradiction was further clarified in 2016 when it was shown, 
using a biorthogonal formalism of \cite{DCB14}, that there is a canonical 
separation of the 
degrees of freedom in a PT-symmetric Hamiltonian into two sets: one 
corresponding to the degrees of freedom associated with its Hermitian 
counterpart Hamiltonian, and one corresponding to the degrees of freedom 
associated with the specification of the metric operator ${\hat g}$ \cite{DCB16}. 
Further, it was shown that for closed quantum systems there are no 
laboratory experiments to adjust or determine the values of the latter 
degrees of freedom. That is, the expectation values of physical observables 
are independent of ${\hat g}$. Thus, perhaps a better characterisation is 
that for a finite-dimensional system the two descriptions are 
\textit{indistinguishable} rather than equivalent. In particular, without the 
ability to adjust the parameters in the operator ${\hat g}$ it is not possible 
to experimentally realise the phase transition discussed above. 

In the infinite-dimensional case, the situation is more complicated even in 
the countable case, because 
the equivalence of eigenvalues between a PT-symmetric Hamiltonian and 
its Hermitian `counterpart' Hamiltonian related by a similarity transformation 
involving the operator ${\hat g}$ is ensured only if both operators ${\hat g}$ 
and ${\hat g}^{-1}$ are bounded. Otherwise, the image of an eigenstate of 
one of the Hamiltonians under the map, or more generally the image of any 
state from the domain of this Hamiltonian, may not belong to the other 
Hilbert space, or to the domain of the other Hamiltonian. 
For many model Hamiltonians studied in the literature, either ${\hat g}$ 
or ${\hat g}^{-1}$ (or both) are unbounded, and in these cases it is not 
simple to determine whether the two descriptions are equivalent: It may be 
the case that there are genuinely new physics of closed quantum systems 
modelled on infinite-dimensional Hilbert spaces, possibly with a partially 
broken PT symmetry, although this remains to be further investigated. 

The above discussion is concerned with closed systems. Around 2006, 
however, it was observed that PT symmetry can be realised in a laboratory 
for open systems by balancing gain and loss \cite{RDM,DC1,KGM}. 
For example, suppose that 
with respect to a choice of reference frame the left side of the system 
absorbs energy, while the right side emits energy, in a symmetric 
configuration, by the same amount. Then under parity (left-right) 
reversal, gain turns 
into loss and loss turns into gain; but time reversal has the same effect so 
that the system is PT symmetric. Intuitively, although the system is open, 
it can behave in a manner similar to a closed one because the overall 
energy is conserved. Such an open system can thus be 
described by a PT-symmetric Hamiltonian, where now the degrees 
freedom associated with the operator ${\hat g}$ can be controlled in a 
laboratory by adjusting, for instance, the gain and loss strengths. It should 
be evident, however, that for such a system to remain in a quasi-equilibrium 
state of energy, the internal energy transport, i.e. the coupling strength of 
the left and the right side of the system, has to be sufficiently strong, for 
otherwise the system cannot redistribute the energy to maintain its 
quasi-equilibrium state. Thus by increasing gain and loss strengths while 
keeping the internal interaction strength fixed, the 
system undergoes a phase transition such that although the Hamiltonian 
remains PT symmetric, its eigenstates are no longer PT symmetric and 
consequently the energy eigenvalues become complex. For PT-symmetric 
systems, such transitions are accompanied by the degeneracies of the 
energy eigenstates, and these critical points are often referred to as 
exceptional points in the literature \cite{Kato,Berry,Heiss,DBEG,Miri}.  

The facts that PT symmetry can be implemented relatively straightforwardly 
for open systems in a laboratory, and that such systems typically exhibit 
nontrivial phase transitions with interesting and sometimes counterintuitive 
features, led many researchers to explore properties of PT-symmetric 
systems both theoretically and experimentally---perhaps most notably in the 
context of optical systems such as optical waveguides. In the context of open 
systems, however, those whose dynamics are described by PT-symmetric 
Hamiltonians are classical rather than quantum mechanical. Likewise, many 
of the experiments that were implemented concern classical systems. The 
reason is because a perfect balancing of gain and loss of energy (or of 
particle number, or of volume) is not feasible in quantum mechanics. 

To see this, it suffices to 
consider the simple system described above where energy 
is absorbed on the left side and emitted on the right side. Suppose that 
$\epsilon$ joules of energy is absorbed on the left side between time $t$ and 
time $t+\rd t$ where $\rd t\ll1$. Then to perfectly balance this, it is necessary 
that exactly $\epsilon$ joules of energy is emitted from the right side precisely 
over the same time interval. Such a process can be achieved classically. 
Quantum mechanically, to realise a perfect PT symmetry one requires a 
sharp precision in both energy and time, but such a precision amounts to 
violating the energy-time uncertainty relation. Another way of seeing the 
issue is to consider a quantum system having a discrete set of energy 
eigenvalues. Then unless the eigenvalues are equally spaced (such as an 
oscillator) and the value of $\epsilon$ coinciding with the energy gap, the 
system will always remain in a state of indefinite energy (i.e. not an energy 
eigenstate) and one can only consider the expectation value of the 
Hamiltonian in order to speak about the energy of the system. 

The impossibility of maintaining a perfect PT symmetry, however, does not 
mean that a PT-symmetric open environment cannot be realised in quantum 
mechanics, because it remains meaningful to speak about balancing gain 
and loss \textit{on average}. A PT-symmetric configuration of the 
absorption and emission \textit{rates} does not violate Heisenberg constraints. 
However, the fact that one cannot determine whether the system has, say, 
absorbed or emitted energy, implies that as time passes, the observer 
gradually loses information about the state of the system. The theoretical 
implication is that it is not possible to characterise the dynamics of a 
PT-symmetric open quantum system in terms of a PT-symmetric Hamiltonian. 
Instead, time evolution is governed by an equation of the Lindblad type, or 
more generally of the Krauss type. 

Focusing in particular on systems described by the Lindblad equation, it 
would be of interest then to enquire what are the characteristic features of 
PT-symmetric open quantum systems and the associated symmetry breaking. 
Indeed, there are growing interests in studying the effects of exceptional 
points in genuinely open quantum systems described by Lindblad equations 
\cite{Hatano,Nori}. With this in mind, 
the purpose here is to explore this question by focusing specifically on the 
behaviour of the speed of the evolution of the state that has been examined 
in \cite{BL}. 

\section{Open quantum system dynamics} 

Intuitively, by an open system $S$ we have in mind one that is coupled to 
another system $B$. The latter may be large enough to act as a 
bath, or may be another small quantum system. Suppose that the total 
system $S+B$ is described by a pure-state wave function 
$|\Psi_{S+B}\rangle$. The state of an open quantum system $S$ is then 
determined by tracing out the degrees of freedom associated with $B$: 
\[ 
{\hat\rho}_S = {\rm tr}_B \left( \frac{|\Psi_{S+B}\rangle\langle\Psi_{S+B}|}
{\langle\Psi_{S+B}|\Psi_{S+B}\rangle} \right) . 
\] 
Provided that the state $|\Psi_{S+B}\rangle$ is not a product state of the 
form $|\Psi_{S+B}\rangle=|\Psi_S\rangle\otimes|\Psi_B\rangle$, or 
equivalently stated if the two systems $S$ and $B$ are entangled, the result 
of the tracing is necessarily a mixed-state density matrix so that we have 
${\hat\rho}_S^2\neq{\hat\rho}_S$. As an example, suppose that $S$ and 
$B$ are both spin-$\frac{1}{2}$ particles, and $|\Psi_{S+B}\rangle$ is the 
pure spin-0 singlet state $\frac{1}{\sqrt{2}}(|\!\uparrow\downarrow\rangle - 
|\!\downarrow\uparrow\rangle)$. Then tracing over $B$ we obtain 
\[ 
{\hat\rho}_S = \half \big( |\!\uparrow\rangle\langle\uparrow\!| + 
|\!\downarrow\rangle\langle\downarrow\!| \big) ,
\]
which is a state distinct from the coherent superposition 
$\frac{1}{\sqrt{2}}( |\!\uparrow\rangle + |\!\downarrow\rangle) $. 

We can think of a density matrix as representing a (classical) statistical 
average over pure states, in the sense that it can be written as the 
expectation of a random pure-state projection operator: 
\[ 
{\hat\rho} = {\mathbb E}\left[ \frac{|\psi\rangle\langle\psi|}
{\langle\psi|\psi\rangle} \right] . 
\]
For instance, suppose that an experimentalist is aware that the state of the 
system will be $|\psi_1\rangle$ if the system absorbs energy, but otherwise 
will be $|\psi_0\rangle$, and that the probability of absorption is $p$. Then 
without a measurement to acquire more information (and thus altering the 
state) one can only assert that the system is in the averaged state 
\[ 
{\hat\rho} = p \frac{|\psi_1\rangle\langle\psi_1|}{\langle\psi_1|\psi_1\rangle}
+(1-p) \frac{|\psi_0\rangle\langle\psi_0|}{\langle\psi_0|\psi_0\rangle} ,
\]
which again is not the same state as the pure state 
$|\psi\rangle=\sqrt{p}\,|\psi_1\rangle + 
\sqrt{1-p}\,|\psi_0\rangle$. Thus the density matrix ${\hat\rho}$ has the 
interpretation of `either $|\psi_0\rangle$ or $|\psi_1\rangle$' in the sense of 
classical probability, whereas the pure state $|\psi\rangle$ has the 
interpretation of `neither $|\psi_0\rangle$ nor $|\psi_1\rangle$' in the sense 
of quantum probability; and the two states are experimentally distinguishable. 
A mixed state density matrix can always be expressed in such an ensemble 
average of pure states. However, given a state ${\hat\rho}$ there are 
uncountably many 
ways of expressing it in terms of averages of pure states, so it is not possible 
to determine which experimental setup has resulted in that state. 

The time evolution of the density matrix that preserves its positivity and 
the trace condition can typically be described by the dynamical equation of 
the Lindblad type: 
\[
\partial_t {\hat\rho} = -\ri[{\hat H},{\hat \rho}] + 
\sum_k \left[ {\hat L}_k {\hat\rho} {\hat L}^\dag_k - \frac{1}{2} \left( 
{\hat L}^\dagger_k {\hat L}_k \rho + {\hat\rho} {\hat L}^\dagger_k {\hat L}_k 
\right) \right] . 
\]
This is the so-called GKLS equation of Gorini, Kossakowski, Sudarshan 
\cite{gorini} and Lindblad \cite{lindblad}, or just the Lindblad equation for short. 
Interestingly, 
in the context of a proposal of Hawking \cite{SWH} that the dynamical equation 
for the density matrix will in general 
not preserve unitarity, Banks, Susskind and Peskin 
\cite{BSP} independently 
explored a general dynamics for the state that preserves positivity and trace 
condition, and arrived at an analogous dynamical equation satisfied by 
the density matrix.  

The Lindblad operators ${\hat L}_k$ characterise the way in which the 
system $S$ interacts with the environment $B$. So for example if there is 
a particle gain at the left side and particle loss at the right side, then the 
dynamics of this system can be characterised by a pair of Lindblad operators 
with ${\hat L}_1 
= \gamma_L {\hat a}_L^\dagger$ and ${\hat L}_2 = \gamma_R 
{\hat a}_R$, expressed in terms of the particle creation and annihilation 
operators, where the parameters $\gamma_L,\gamma_R$ determine the 
gain/loss strengths. That an averaged balancing of gain and loss can be 
realised in this way to describe a PT-symmetric quantum system in the 
context of Bose-Einstein condensates has been known for long to the 
Stuttgart group \cite{Wunner1,Wunner2}. In the case of a many body system, 
such a system can be treated effectively in terms of a mean-field 
approximation, which in turn may be described by an effective 
PT-symmetric Hamiltonian \cite{Graefe1}. 

From the linearity of the dynamics we can think of the right side of the 
Lindblad equation as representing the action of a Liouville operator 
${\cal L}$ on ${\hat\rho}$, and write 
\[ 
\partial_t {\hat\rho}={\cal L}{\hat\rho} . 
\] 
Here the Liouville operator ${\cal L}$ can be interpreted as a superoperator, 
which becomes evident if we express operators using the index notation by 
writing $\partial_t \rho^\alpha_\beta = {\cal L}^{\alpha\alpha'}_{\beta\beta'} 
\rho^{\beta'}_{\alpha'}$ with repeated indices summed over. It turns out that 
the eigenvalues of ${\cal L}$ are real 
or else come in complex conjugate pairs, just like any PT-symmetric 
operator. Thus, from a certain point of view one could argue that every open 
system dynamics is PT symmetric, although exactly what that might mean 
physically is not well understood.

\section{Unitary evolution speed for quantum states} 

There are various features of the dynamics of an open system that one can 
investigate, but following \cite{BL}, 
here we shall be focused on one particular aspect concerning 
the speed of evolution of the state. The evolution speed of a quantum system 
in itself is of interest for several reasons. For example, to implement a quantum 
operation such as a quantum algorithm in a quantum computer, it is useful 
to have an idea on how fast a given task can be realised. The evolution speed 
is also linked to the sensitivity or stability of the state against changes in time, 
which in turn is useful in arriving at error lower bounds in estimating quantum 
states \cite{BH0}. 

For pure states, the evolution speed under a unitary motion is given by the 
Anandan-Aharonov relation \cite{AA}. 
The idea can be described as follows. We start 
with the Schr\"odinger-Kibble equation 
\[ 
\frac{\rd}{\rd t}\, |{\psi}(t)\rangle = -\ri\frac{1}{\hbar} \left( {\hat H} - 
\langle\psi(t)|{\hat H}|\psi(t)\rangle\right) |\psi(t)\rangle 
\] 
for the dynamics of the state. Note that the Schr\"odinger-Kibble equation is 
a nonlinear differential equation defined on Hilbert space. However, because 
the energy expectation $E=\langle\psi(t)|{\hat H}|\psi(t)\rangle$ is a constant 
of motion, the equation is indeed linear along each unitary orbit. The 
effect of removing the energy expectation from the Hamiltonian is to eliminate 
dynamical phase under the evolution, and the equation has the advantage 
that for a stationary state we have $|{\dot\psi}(t)\rangle=0$ and hence we 
automatically 
recover the time-independent Schr\"odinger equation ${\hat H}|\psi\rangle = 
E|\psi\rangle$, which in undergraduate quantum mechanics texts can only be 
derived by evoking the somewhat mysterious correspondence between 
energy and time derivative. Mathematically, the Schr\"odinger-Kibble equation 
gives what is called a horizontal lift of the Schr\"odinger equation in Hilbert 
space to the projective Hilbert space where the tangent vector $|\dot\psi\rangle$ 
is everywhere orthogonal to $|\psi\rangle$. 

If the state $|\psi\rangle$ of a system satisfies the Schr\"odinger-Kibble 
equation, then the squared speed of evolution is determined by 
\[ 
v^2(t) = 4 \langle {\dot\psi}(t)|{\dot\psi}(t)\rangle,  
\] 
where the factor of $4$ here is conventional in fixing the scale of the metric, 
and is related to the fact that the Bloch sphere of spin-$\frac{1}{2}$ systems 
has radius $\frac{1}{2}$. 
Alternatively, if we work with the solutions to the conventional Schr\"odinger 
equation $|{\dot\psi}(t)\rangle=-\ri\hbar^{-1}{\hat H}|\psi(t)\rangle$, then we 
can define the proper `velocity' vector by 
\[
|v(t)\rangle = |{\dot\psi}(t)\rangle - 
\langle{\psi}(t)|{\dot\psi}(t)\rangle\,  |\psi(t)\rangle , 
\]
and calculate $v^2(t)=4\langle v(t)|v(t)\rangle$. 
Either way, we deduce as a result the Anandan-Aharanov relation: 
\[
v^2(t) = \frac{4 \Delta H^2}{\hbar^2} , 
\]
which asserts that the speed of the evolution is determined by the energy 
uncertainty in the state. But because energy uncertainty is constant of 
motion under unitarity, we see that the speed $v(t)$ only depends on the 
initial state, and is independent of $t$.

For mixed states the situation is a little more complicated. The reason is 
because the notion of speed, which is distance divided by time, depends on 
the choices of distance and time. As for time, this is measured by the 
parameter $t$ in the evolution equation, but we are still left with the choice 
of distance. For pure states we are able to circumvent the discussion because 
there is only one unitary invariance notion of distance, given by the 
Fubini-Study metric on the state space \cite{BH1}. For mixed states, there 
are choices in the space to represent the density matrix (for example, a 
Hilbert space or a Euclidean space), and these different choices result in 
different distance measures. 

To make a direct comparison with the speed for pure states evolution in 
Hilbert space, we consider here embedding of density matrices in a Hilbert 
space. This is because while the density matrix itself has trace unity, in infinite 
dimensions its square need not have finite trace, if it has continuous 
spectrum. So instead of working with ${\hat\rho}$ we consider 
a Hermitian square-root of the density matrix given by 
\[
{\hat\rho} \to {\hat\xi} = \surd{\hat\rho} 
\]
so that ${\hat\rho}={\hat\xi}^2$. The choice of ${\hat\xi}$ is not unique, but 
any such square-root density matrix would suffice. Then integrability of 
${\hat\rho}$ makes ${\hat\xi}$ square-integrable, i.e. it belongs to a real Hilbert 
space endowed with the trace norm. Under a unitary time evolution, the 
dynamical equation satisfied by ${\hat\xi}$ is 
\[
\partial_t {\hat\xi} = -\ri[{\hat H},{\hat \xi}] .
\]
Then the squared speed of evolution is defined by 
\[
v^2(t) =  {\rm tr}(\partial_t{\hat\xi}\, \partial_t{\hat\xi}) , 
\]
and a short calculation shows that 
\[
v^2(t) = 2 \left[ {\rm tr}({\hat H}^2{\hat\rho}) 
- {\rm tr}({\hat H} \surd{\hat\rho} {\hat H}\surd{\hat\rho}) \right] . 
\]
The right side is twice the Wigner-Yanase skew information measure 
\cite{Luo,DCB0}. It is not difficult to show that 
\[ 
\Delta H^2 \geq 
{\rm tr}({\hat H}^2{\hat\rho}) 
- {\rm tr}({\hat H} \surd{\hat\rho} {\hat H}\surd{\hat\rho}) \geq 0 , 
\] 
where the upper bound $\Delta H^2$ is attained for a pure state, 
because if ${\hat\rho}=|\phi\rangle\langle\phi|$ for some $|\phi\rangle$ then 
$\surd{\hat\rho}=|\phi\rangle\langle\phi|$
so ${\rm tr}({\hat H} \surd{\hat\rho} {\hat H}\surd{\hat\rho})=\langle\phi|
{\hat H}|\phi\rangle^2$; whereas the lower bound $0$ is attained if 
${\hat\rho}$ is a stationary state so that $[{\hat\rho},{\hat H}]=0$. It follows 
that the speed of evolution under a unitary motion is somewhat slowed 
down for mixed states, as compared to their pure-state counterparts. It is 
worth noting that the energy variance $\Delta H^2$ is nonzero for all impure 
stationary states, whereas the Wigner-Yanase skew information vanishes 
for all stationary states. It follows that the Anandan-Aharonov relation for 
the speed of unitary evolution is valid only for pure states; for general 
states -- pure or mixed -- the speed is given by the Wigner-Yanase skew information, and this is related to the fact that the Fisher information 
associated with time estimation for mixed states is not given by the 
energy uncertainty \cite{DCB0}.

\section{Embedding mixed states in Euclidean space} 

In the case of an open system dynamics, to make a direct comparison to 
the foregoing analysis on unitary motions it would be desirable to first 
formulate the evolution equation satisfied by the square-root density matrix 
${\hat\xi}$ and then work out the squared-speed by the formula 
$v^2(t) = 2\, {\rm tr}(\partial_t{\hat\xi}\, \partial_t{\hat\xi})$. However, the 
evolution equation for ${\hat\xi}$ associated with a Lindblad dynamics for 
${\hat\rho}={\hat\xi}^2$ is not known. Therefore, rather than working in a 
Hilbert space and useing a Hilbert-space norm to determine the speed, we 
shall instead regard the space of density matrices as a subspace of a 
Euclidean space and use the Euclidean norm to evaluate the speed 
\cite{BL,CPBM,CPBM2}. It 
follows that in the unitary limit we will not recover the Wigner-Yanase 
skew information because a different distance measure is used. Indeed 
there are various different measures being considered in the literature to 
analyse the speed of evolution 
\cite{Pati,DC,AMR,Funo,Raam,Deffner,Taddei,Plenio}. 

The space of density matrices in a Hilbert space ${\cal H}^n$ of dimension 
$n$ forms a subset of the interior of a sphere $S^{n^2-2}$ in a Euclidean 
space ${\mathds R}^{n^2-1}$. To see this, recall that the outer boundary of 
the space of density matrices consists of pure states. Then writing 
$|\phi\rangle=(\phi^1,\phi^2,\ldots,\phi^n)$ for a normalised pure state vector 
we can define coordinates of pure states in ${\mathds R}^{n^2}$ by 
$(x^k, x^{hk}, y^{hk})$, where $h,k=1,2,\ldots,n$, $k\neq h$, if we set 
\[ 
x^k = \sqrt{2}\, \phi^k {\bar\phi}^k , \quad x^{hk} = \phi^h {\bar\phi}^k + 
\phi^k {\bar\phi}^h, \quad y^{hk} = \ri( \phi^h {\bar\phi}^k - 
\phi^k {\bar\phi}^h ) . 
\]
Then the trace condition on $|\phi\rangle\langle\phi|$ implies that we have 
a linear constraint $x^1+x^2+\cdots+x^n=\sqrt{2}$ corresponding to a 
hyperplane ${\mathds R}^{n^2-1}$, and hence the pure states lie on a 
sphere $S^{n^2-2}$, forming the boundary for general mixed states \cite{DCB1}. 
Accordingly, every density 
matrix ${\hat\rho}$ can be thought of as being represented by a vector 
${\boldsymbol r} \in {\mathds R}^{n^2-1}$. 

We let $\{{\hat \sigma}_j\}_{j=0,\ldots,n^2-1}$ be an orthonormal basis 
for the linear space of bounded operators on ${\cal H}^n$ with 
the Hilbert-Schmidt inner product 
\[
\langle {\hat\sigma},{\hat\tau}\rangle
={\rm tr}\left({\hat \sigma}^\dagger{\hat \tau}\right) , 
\] 
and set ${\hat \sigma}_0=n^{-1/2}{\mathds 1}$. Hence the 
operators $\{{\hat \sigma}_j\}_{j=1,\ldots,n^2-1}$ are trace free, and 
together they satisfy the orthonormality condition 
$$\langle {\hat \sigma}_i,{\hat \sigma}_j\rangle 
= \delta_{ij}.$$  
An arbitrary density matrix ${\hat\rho}$ can then be expressed in the form 
\[
{\hat\rho} = \frac{1}{\sqrt{n}}\, {\hat \sigma}_0 
+ \sum_{j=1}^{n^2-1} r_j \, {\hat \sigma}_j, 
\]
where $r_j={\rm tr}({\hat\rho}{\hat \sigma}_j)$. For a pure state we have 
\[
{\rm tr}({\hat\rho}^2) = \frac{1}{n} + \sum_{j=1}^{n^2-1} r_j^2 = 1,
\]
from which it follows that the squared radius of the sphere $S^{n^2-2}$ in 
${\mathds R}^{n^2-1}$ is given by $1-n^{-1}$. For $n=2$ we then get the 
usual Bloch sphere with radius $\frac{1}{2}$. 

Consider now a one-parameter family of density matrices ${\hat\rho}(t)$ 
satisfying the Lindblad equation 
\[
\partial_t {\hat\rho} = -\ri[{\hat H},{\hat \rho}] + 
\sum_k \left[ {\hat L}_k {\hat\rho} {\hat L}^\dag_k - \frac{1}{2} \left( 
{\hat L}^\dagger_k {\hat L}_k \rho + {\hat\rho} {\hat L}^\dagger_k {\hat L}_k 
\right) \right] , 
\]
along with an initial condition ${\hat\rho}(0)$. Substituting 
\[
{\hat\rho} = \frac{1}{\sqrt{n}}\, {\hat \sigma}_0 
+ \sum_{j=1}^{n^2-1} r_j \, {\hat \sigma}_j 
\]
in here, we find that the dynamical equation in ${\mathds R}^{n^2-1}$ is given by
\[
{\dot r}_i = \sum_{j=1}^{n^2-1} \Lambda_{ij} r_j + b_i , 
\]
where 
\[
\Lambda_{ij} = \, {\rm tr} \left[ -\ri\, [{\hat\sigma}_j,{\hat\sigma}_i] 
{\hat H} + \sum_k {\hat L}_k{\hat\sigma}_j{\hat L}_k^\dagger {\hat\sigma}_i
- \half \sum_k \left( 
{\hat L}^\dagger_k {\hat L}_k {\hat\sigma}_j {\hat\sigma}_i + 
{\hat L}^\dagger_k {\hat L}_k {\hat\sigma}_i {\hat\sigma}_j \right) \right] 
\]
is a real matrix, and 
\[
b_i = \frac{1}{n} \sum_k {\rm tr} \left( [{\hat L}_k,{\hat L}^\dagger_k ] 
{\hat\sigma}_i \right) 
\]
is a real vector. Writing the dynamical equation in the form 
$\partial_t {\hat\rho}={\cal L}{\hat\rho}$, 
the components ${\cal L}_{ij}$ of ${\cal L}$ in the basis 
$\{{\hat\sigma}_j\}$ are 
\[
{\cal L}_{ij} =  \langle {\hat\sigma}_i, {\cal L}{\hat\sigma}_j\rangle  = 
{\rm tr}\left( {\hat\sigma}_i {\cal L}{\hat\sigma}_j \right),
\] 
and we have 
$({\cal L}{\hat\rho})_j = \sum_i {\rm tr}( {\hat\sigma}_j {\cal L}{\hat\sigma}_i) 
\, {\rm tr} ( {\hat\sigma}_i {\hat\rho})$. 
Therefore, writing 
\[
{\dot r}_j = \sum_{i=1}^{n^2-1} {\cal L}_{ji} r_i + \frac{1}{\sqrt{n}}\,{\cal L}_{j0}
\, ,
\]
we deduce that  
${\cal L}_{ji} = \Lambda_{ji}$ for $i,j\neq0$ and ${\cal L}_{j0} = \sqrt{n}\,b_j$. 
Because the real matrix ${\cal L}_{ji}$ is in general not symmetric, its 
eigenvalues are 
either real or else come in complex conjugate pairs. If the Lindblad operators 
are normal, then we have $b_j=0$ and the eigenvalues of the Liouville 
operator are thus determined by the matrix $\Lambda_{ji}$.

\section{Evolution speed for mixed states} 

In terms of the Euclidean norm, the squared speed of evolution of the state 
in ${\boldsymbol r} \in {\mathds R}^{n^2-1}$ is given by 
\[
v^2(t) = \sum_{j=1}^{n^2-1} {\dot r}_j^2 = {\rm tr} \left[ ({\cal L}{\hat\rho})^2\right]. 
\]
Writing 
\[
\mathcal{L}{\hat\rho} = -\ri [{\hat H},{\hat\rho}] + \mathcal{D}{\hat\rho} 
\] 
for the Lindblad equation, where 
\[
\mathcal{D}{\hat\rho} = \sum_k \left[ {\hat L}_k {\hat\rho} {\hat L}^\dag_k - \frac{1}{2} 
\left( {\hat L}^\dagger_k {\hat L}_k \rho + {\hat\rho} {\hat L}^\dagger_k {\hat L}_k 
\right) \right] ,
\]
we obtain: 
\[
v^2(t) = 2\left[{\rm tr}\left({\hat H}^2{\hat\rho}^2\right) - {\rm tr}\left( {\hat H} 
{\hat\rho} {\hat H} {\hat\rho}\right)\right] 
- 2\ri\, {\rm tr}\left( {\hat\rho}\, [{\cal D}{\hat\rho},{\hat H}]\right) + {\rm tr} \left[\left(\mathcal{D}{\hat\rho}\right)^2\right].
\]
We see that there are three terms contributing to the speed of evolution. 
The contribution to $v^2$ from the unitary evolution resembles, but is 
different from, the Wigner-Yanase skew information $I = {\rm tr}( {\hat H}^2{\hat\rho}) - {\rm tr}({\hat H}{\sqrt{\hat\rho}}{\hat H}{\sqrt{\hat\rho}})$. 
Let us call 
\[ 
S(X)={\rm tr}(
{\hat X}^\dagger{\hat X}{\hat\rho}^2) - 
{\rm tr}({\hat X}{\hat\rho}{\hat X}^\dagger
{\hat\rho}) 
\]
the `modified skew information' for ${\hat X}$, which 
reduces to $\Delta X^2=\langle{\hat X}^\dagger{\hat X}\rangle - 
\langle{\hat X}^\dagger\rangle\langle{\hat X}\rangle$ for a pure state \cite{BL}. 
In particular, if $\hat X$ is Hermitian then $S(X)$ is just its variance. Thus for 
pure states we recover the Anandan-Aharonov relation. This follows on account 
of the fact that the metric on the space of pure states induced by the ambient 
Euclidean metric on ${\mathds R}^{n^2-1}$ is the Fubini-Study metric 
\cite{DCB1}. 

In contrast to unitary time evolution, in an open system the velocity will in 
general obtain a `radial' component so that the purity ${\rm tr}({\hat\rho}^2)$ 
changes. The measure of purity-change rate is then given by the squared 
magnitude of the radial velocity
\[
v^2_R(t) = \frac{({\boldsymbol r}\cdot {\dot{\boldsymbol r}})^2}{
{\boldsymbol r}\cdot {\boldsymbol r}} = \frac{[{\rm tr}({\hat\rho} \mathcal{L}
{\hat\rho})]^2}{{\rm tr}[({\hat\rho}-n^{-1}{\mathds 1})^2]},
\] 
which vanishes for unitary dynamics. A short calculation shows remarkably 
that  
\[
v_R(t) = \sum_k \frac{S(L_k)}{\sqrt{{\rm tr} \left[({\hat\rho}-n^{-1}{\mathds 1})^2\right]}} . 
\]
In other words, the speed of the change of the purity is determined by 
the modified skew information associated with the Lindblad operators. 
Note that if for some $t$ we have ${\hat\rho}(t)=n^{-1}{\mathds 1}$, 
the state of total ignorance, the denominator for $v_R(t)$ vanishes but 
in this state we also have $S(L)=0$ for all ${\hat L}$ so that $v_R(t)$ 
remains finite. 

Let us examine the behaviour of the evolution speed via illustrative 
examples of $2\times2$ models \cite{BL}. (i) For the first example we take the 
Hamiltonian to be ${\hat H}=\frac{1}{2}g {\hat\sigma}_z$ and the Lindblad operator to be ${\hat L}=\sqrt{\gamma}{\hat\sigma}_z$. Because the 
Hamiltonian and the Lindblad operator commute here, we get a pure 
decoherence dynamics in which the off-diagonal elements of the density 
matrix decay exponentially, while the diagonal elements remain constants 
of motion. In this example it is easy to deduce that 
\[
v^2(t) = \re^{-4\gamma t}(4\gamma^2 + g^2)[r^2_x(0) + r^2_y(0)],
\]  
and hence that $v(t)$ decreases exponentially in time. The motion of the 
state under the dynamics can easily be visualised. Suppose that the initial 
state is a pure state on the surface of the Bloch sphere in ${\mathds R}^3$. 
Then under the Lindblad dynamics, the state spirals inwards in the $x-y$ 
plane while keeping the value of $z={\rm tr}({\hat\sigma}_z{\hat\rho})$ 
constant, and is eventually absorbed to the $z$-axis. 

(ii) In the next example we let the Hamiltonian be ${\hat H}=
\frac{1}{2}g{\hat\sigma}_x$, and we choose the Lindblad operator to be 
${\hat L}=\sqrt{\gamma}{\hat\sigma}_z$ so that it does not commute with 
${\hat H}$. One can think of this example as the simplest nontrivial model of 
a PT-symmetric open quantum system. It is important to keep in mind, however, 
that the notion of PT-symmetry in quantum mechanics for open systems is 
distinct from that in classical systems. 
Suppose that one starts with a pure initial state. 
Then the effect of the Lindblad operator ${\hat L}=
\sqrt{\gamma}{\hat\sigma}_z$ on the space of pure states is 
to turn the state into an eigenstate of ${\hat\sigma}_z$. That is, an initial 
pure state $|\psi\rangle$ turns into either $|\!\uparrow_z\rangle$ or 
$|\!\downarrow_z\rangle$, in a manner similar to the effect of a 
measurement being performed on the observable ${\hat\sigma}_z$. 
Thinking of 
${\hat\sigma}_z$ as representing `energy' we see that sometimes the 
system will gain energy so that the state of the system converges to the 
eigenstate of ${\hat\sigma}_z$ with larger eigenvalue; sometimes it will lose 
energy so that the state of the system converges to the other eigenstate of 
${\hat\sigma}_z$ with smaller eigenvalue---but which way it might go is 
random and is governed by the Born probability rule. This picture is often 
referred 
to as the stochastic unravelling of the Lindblad equation: The pure state 
evolves randomly, but the expectation of the random pure-state projector 
gives the state (density matrix) of the system that obeys a deterministic 
equation of the Lindblad type. The important point 
is that while each realisation of such a process is random, on average the 
energy is conserved in the sense that ${\rm tr}({\hat\sigma}_z{\hat\rho})$ 
is constant of motion under the pure Lindblad dynamics (when ${\hat H}
=0$). Hence on average the gain and loss are balanced, although in each 
realisation the system either gains energy or loses energy. This is in 
contrast to classical situations whereby (in the unbroken phase) gain 
and loss are exactly balanced; not merely on average. 

The addition of the unitary term generated by a Hamiltonian that does not 
commute with the Lindblad operator, however, breaks this balance and the 
system no longer has conserved (on average) observables. Nevertheless, 
the characteristic features of a PT-symmetric system described by the 
Hamiltonian ${\hat H}=g{\hat\sigma}_x-\ri\gamma{\hat\sigma}_z$, and in 
particular the impact of the exceptional points $g=\pm\gamma$ for this 
Hamiltonian, manifests 
itself. To see this we note that the matrix elements $\Lambda_{ij}$ of the 
Liouville operator can easily be worked out to give: 
\[
\Lambda = \begin{pmatrix} -2\gamma & 0 & 0\\0 &-2\gamma & -g\\0 
& g & 0 \end{pmatrix} 
\]
along with $b_i=0$. Hence the four eigenvalues of the Liouville operator 
are $0$, $-2\gamma$, and $-\gamma \pm \sqrt{\gamma^2-g^2}$. It follows 
that in the region of unbroken PT-symmetry $g>\gamma$ the eigenvalues 
are either real or come in complex conjugate pairs;  at the exceptional point 
$g=\gamma$ the PT-symmetry gets broken; and in the broken phase 
$g<\gamma$ all the eigenvalues are real. At first this may appear to be the 
reversal of the usual situation with PT symmetry, but recall that we have 
defined the Liouville operator according to the equation 
$\partial_t {\hat\rho}={\cal L}{\hat\rho}$ without the complex number. 

We remark 
that the lack of real eigenvalues for $\ri{\cal L}$ is due to the fact that 
there is always loss of information in an open quantum dynamics, leading 
to an overall decay; but the signature of PT symmetry is seen in the 
remaining details of the dynamics. Thus, from the existence of PT-symmetry 
breaking in this model, we expect the system to 
exhibit different behaviours in each of these phases. Indeed, the solutions 
to the dynamical equations are: 
\begin{align} 
& \re^{\gamma t}\, r_x(t)=\re^{-\gamma t} \, r_x(0), \nonumber \\ 
& \re^{\gamma t}\, r_y(t) = \left[\cos \omega t - (\gamma/\omega) 
\sin\omega t\right]r_y(0) - \left[(g/\omega) \sin \omega t\right] r_z(0), 
\nonumber \\ 
& \re^{\gamma t}\, r_z(t) =  \left[(g/\omega) 
\sin \omega t \right] r_y(0) + \left[\cos \omega t + 
(\gamma/\omega)\sin \omega t\right] r_z(0), \nonumber 
\end{align}
with $\omega = \sqrt{g^2-\gamma^2}$. We see therefore that the solutions 
are oscillatory in the unbroken phase $g>\gamma$, whereas in the broken 
phase $g<\gamma$ the oscillations are completely suppressed. Notice 
that by writing the solutions in the form $\re^{\gamma t}\, r_\alpha(t)$, 
$\alpha=x,y,z$, we remove the background damping effect so that the effect 
of PT symmetry becomes transparent. 

The squared speed of evolution $v^2(t)$, its radial component $v_R^2(t)$, 
and its tangential component $v^2_T(t) = v^2(t) - v^2_R(t)$, are shown 
below for a system prepared initially in the spin-$z$ up state 
$|\psi(0)\rangle  = |\!\!\uparrow\rangle$. Because the initial state is chosen to 
be an eigenstate of ${\hat L}$, we have $v_R(t)=0$ at $t=0$. In the unbroken 
phase, the speed exhibits a decay superimposed with oscillations. Here 
$v_R(t)$ oscillates periodically with the period $\tau=\pi/\sqrt{g^2-\gamma^2}$, 
where the minima correspond to the times at which the Bloch vector is aligned 
with the $z$-axis, i.e. when ${\cal D}\hat\rho\propto{\hat\rho}$. 
Moving into the broken phase, the speed decays rapidly at short times and 
the oscillation in $v_T(t)$ is completely damped out. However, in this phase 
the velocity remains nonzero for a longer duration. 

\begin{figure}[t]
      \centering
      %\vspace{-9.0cm}
        \includegraphics[width=0.31\textwidth]{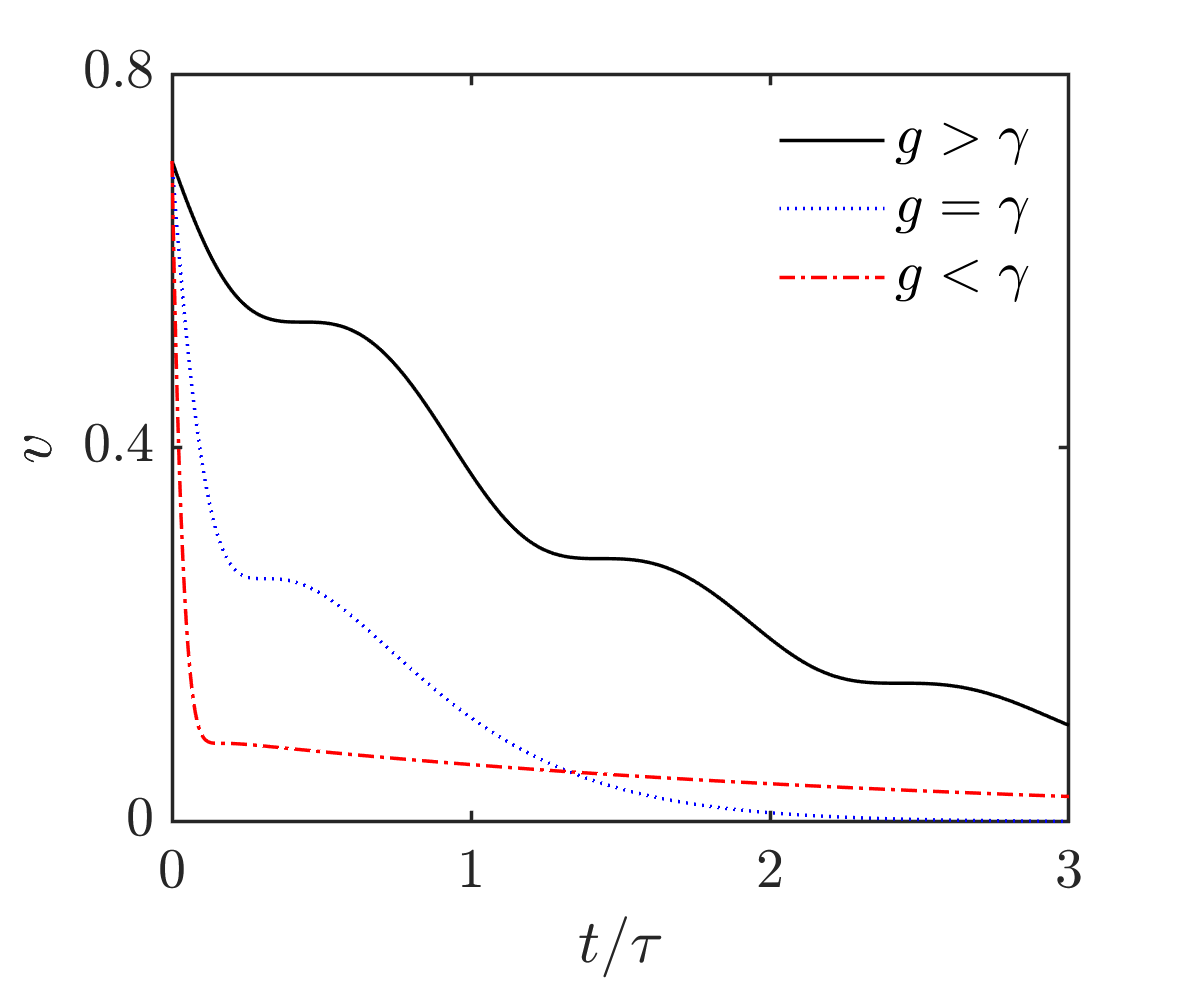} 
%        \end{figure}
%\begin{figure}[t]
%	\centering
	%\vspace{-9.0cm}
        \includegraphics[width=0.31\textwidth]{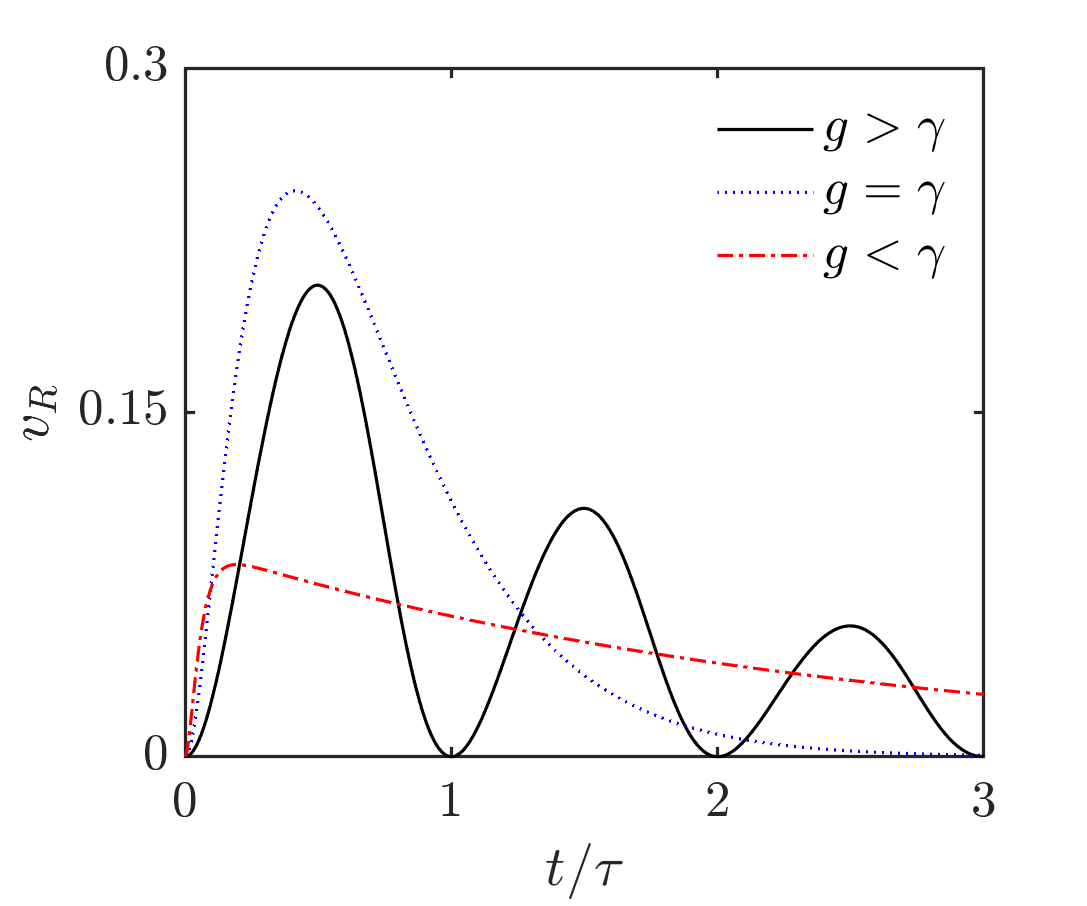}%\hfill 
%\end{figure}
%\begin{figure}[t]
%	\centering
	%\vspace{-9.0cm}
        \includegraphics[width=0.31\textwidth]{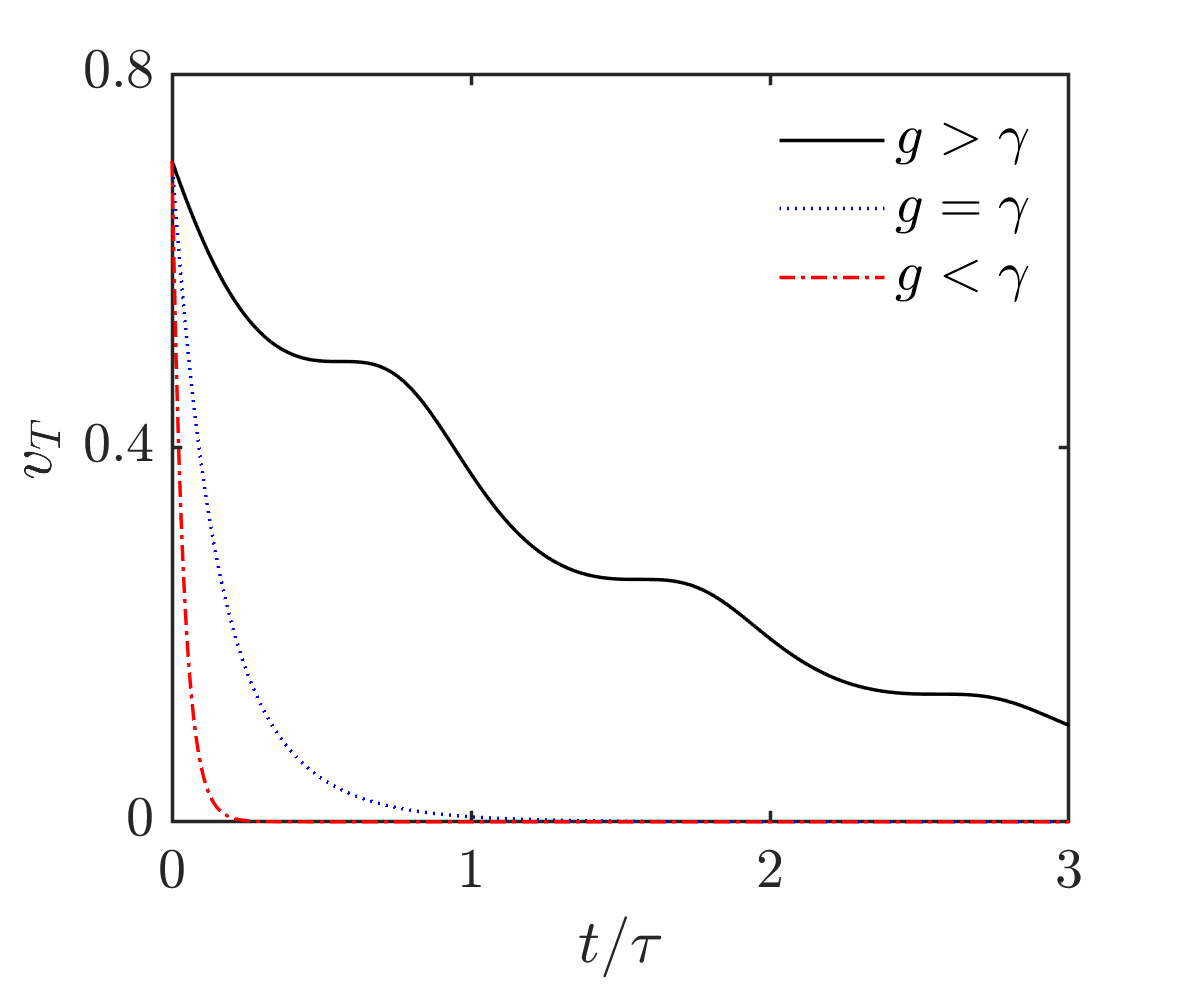}
        \caption{The behaviour of the evolution speed $v(t)$, its radial 
        component $v_R(t)$, and its tangential component $v_T(t)$ are 
        plotted here in the PT-symmetric model for which 
        ${\hat H}=\frac{1}{2}g{\hat\sigma}_x$ and ${\hat L}=
        \sqrt{\gamma}{\hat\sigma}_z$. Each plot shows the speed as a 
        function of time in three phases: the unbroken phase $g>\gamma$, 
        the critical point $g=\gamma$, and the broken phase $g<\gamma$.} 
                \end{figure}
               
\vspace{0.3cm}
\begin{footnotesize}
\noindent {\bf Acknowledgements}. The author 
acknowledges support from the Russian Science 
Foundation, grant 20-11-20226, and is grateful to B. Longstaff 
for discussion. 
\end{footnotesize}

\end{document}